\documentclass[12pt]{iopart}

\usepackage{graphicx} 
\usepackage{amssymb}
\usepackage{iopams} 
\newcommand {\be}{\begin{equation}}
\newcommand {\ee}{\end{equation}}
\newcommand {\ba}{\begin{eqnarray}}
\newcommand {\ea}{\end{eqnarray}}

\begin{document}

\title[ ]{Probing Hamiltonian dynamics by means of the 0-1 test for chaos}

\author{M Romero-Bastida$^1$, M A Olivares-Robles $^1$ and E Braun $^2$}
\address{$^1$ SEPI-ESIME Culhuac\'an, Instituto Polit\'ecnico Nacional, Av. Santa Ana No. 1000, Col. San Francisco Culhuac\'an, Delegaci\'on Coyoac\'an, Distrito Federal 04430, M\'exico}
\address{$^2$ Departamento de F\'\i sica, Universidad Aut\'onoma Metropolitana Iztapalapa, Apartado Postal 55--534, M\'exico, Distrito Federal 09340, M\'exico}

\ead{mromerob@ipn.mx}
\date{\today} 

\begin{abstract}

A recently proposed test for chaos [Gottwald G A and Melbourne I 2004 \textit{Proc. R. Soc. London A} {\bf 460} 603-611] is employed to probe the Hamiltonian dynamics of a one-dimensional anharmonic oscillator lattice. For a homogeneous (uniform mass) lattice in the weakly chaotic regime and for a heavy impurity embedded in the lattice, the results stemming from the time record of the position and momentum of a single oscillator in the former case, and for that same variables corresponding to the impurity in the latter, are inconclusive to determine the dynamical regime of the system. This seemingly odd behavior has its origin in the insufficient time series length employed. Nevertheless, for both cases the necessary time record length needed to obtain the correct result renders the test impractical. In particular, for the second case, specially in the large system size limit (which is the physically relevant one due to its connection with Brownian motion), the estimated length of the position time series required by the test to correctly classify the signal is beyond the reach of present-day computer capability. Thus our results indicate that the proposed test, for the aforementioned cases of Hamiltonian chaos, affords no clear advantage over conventional phase space reconstruction methods.
\end{abstract}

\pacs{05.45.Tp, 05.45.Jn, 05.45.Pq, 05.40.Jc}
\submitto{\JPA}


\section{Introduction\label{sec:Intro}}

A long-standing fundamental issue in the theory of time series analysis is to determine whether a complex time series is regular, deterministically chaotic, or random. Recently, a test, termed 0-1 test, for distinguishing regular from chaotic dynamics in deterministic dynamical systems has been proposed~\cite{Test1}. The input is the time series of a relevant variable and the output is $0$ or $1$, depending on whether the dynamics is regular or chaotic, respectively. The test has been applied successfully to the H\'enon-Heiles and Lorenz systems, being found useful as a marker of the transition from regularity to chaos~\cite{Barrow} and of the transition between quasiperiodic dynamics and a strange nonchaotic attractor~\cite{Dawes}. Positive results of its application to simple experimental time series have also been reported~\cite{SIADS}. The aforementioned results seem to support the claim made in~\cite{Test1} that the dimension of the dynamical system and the form of the underlying equations are irrelevant, since the 0-1 test does not require the phase space reconstruction of conventional nonlinear time-series methods~\cite{Kantz}. However, mainly from the analysis of the logistic map previously studied in~\cite{Test1}, it was also claimed that the 0-1 test is not useful for exploratory purposes, specially for the analysis of data with little \emph{a priori} knowledge of the underlying dynamics~\cite{Jing05}. For this particular system these assertions later on proved to be largely unjustified since they stemmed from a misapplication of the test~\cite{Reliability}. Therefore, it is still important to continue exploring the range of applicability, and possible hitherto unacknowledged limitations, of the 0-1 test in order to avoid such misinterpretations.

The available evidence is consistent with the claim that, as long as the system under study is truly deterministic, the 0-1 test is valid. A system that fulfills the aforementioned condition and successfully studied by means of the 0-1 test in~\cite{Test1} is that described by the driven and damped Kortweg-de Vries (KdV) equation. Notwithstanding it formally describes an infinite-dimensional dynamical system, it is now well accepted that the chaotic solutions of such systems evolve in an effective manifold (attractor) of finite Hausdorff dimension~\cite{Robinson}. Therefore, further research is needed to support the claim that the 0-1 test effectively characterizes the dynamics of high-dimensional systems~\cite{Test1}. The choice of the most adequate model system wherewith to explore the validity of the aforementioned claim is afforded by the observation that the KdV equation is an integrable approximation to a discrete many-degrees-of-freedom dynamical system, namely the Fermi-Pasta-Ulam (FPU) anharmonic oscillator lattice~\cite{Chirikov,Ford,Ponno}.

In this work we apply this technique to artificial (computer-generated) time series derived from the position and momentum of a single oscillator in a FPU lattice to asses its usefulness in identifying signals of unmistakable dynamical origin to establish if the 0-1 test can be applied with enough confidence to signals for which no \emph{a priori} information of its dynamical origin, either regular or chaotic, is available. Further insights about the range of applicability can be obtained if structural changes are introduced in the lattice, i.e. if a heavy impurity is coupled to the oscillator chain. The importance of this modification stems from the fact that a heavy particle coupled to a many-degrees-of-freedom system, under conditions independent of the regular or chaotic character of the latter, performs Brownian motion (See~\cite{RomeroBraun} and references therein). Now, some time ago it was proposed that, from the time record of the position of a Brownian particle (BP), which is an experimentally accessible variable, it could be possible to detect the chaotic dynamics of the fluid wherein the particle is suspended by means of nonlinear time-series methods~\cite{Gaspard}. However, it was shown later on that such methods render inconclusive results for both microscopic chaos detection~\cite{Dettmann} and randomness vs chaos distinction~\cite{Cencini}. Since a colloidal particle embedded in a fluid is, in principle, a deterministic dynamical system governed by Newton's equations of motion, the 0-1 test can be applied to the experimental records of the position of a BP, as well as to the computer-generated position time series of the heavy impurity in the large system size limit of the FPU lattice wherein it is embedded. The proposed application affords a novel way to corroborate the claim made in~\cite{Reliability} that this technique avoids certain well-documented drawbacks of conventional nonlinear time series methods~\cite{Kantz}.

This paper is organized as follows. In Sec.\ \ref{sec:Test} we briefly review the 0-1 test and the relevant details of its implementation. Sec.\ \ref{sec:Model} describes the employed model and the relevant details of its numerical integration. Secs.\ \ref{sec:IC} and\ \ref{sec:WCR} present the results for two different dynamical regimes of the FPU model. In Sec.\ \ref{sec:TSS} we investigate the time scale separation effect induced by embedding a heavy impurity in the lattice. The large system size limit of the variable that conveys the result of the 0-1 test obtained from the position time series of the heavy impurity and the comparison with the results obtained from an experimental record of the position of a BP are given in Sec.\ \ref{sec:BM}. In Sec.\ \ref{sec:Disc} we discuss the previous results and present our conclusions.

\section{The 0-1 test for chaos\label{sec:Test}}

The employed method starts with a finite data set $\{\phi(t_{\alpha})\}_{\alpha=1}^{\cal N}$ sampled at discrete times $t_{\alpha}\equiv\alpha\tau$, with sampling time $\tau$. Here $\phi(t_{\alpha})$ is a one-dimensional observable obtained from the underlying dynamics. First, for a given $c\in\Re$, define
\be
\eqalign{\xi(t_{\alpha})=\sum_{j=1}^{\alpha}\phi(t_j)\cos(jc)   \cr
\eta(t_{\alpha})=\sum_{j=1}^{\alpha}\phi(t_j)\sin(jc),}\label{xidef}
\ee
where $\alpha=1,2,3,\ldots$ Next, for one of the above variables, say $\xi(t_{\alpha})$, the mean square displacement is computed as
\be
M(t_{\alpha})=\lim_{{\cal N}\rightarrow\infty}{1\over{\cal N}-\alpha}\sum_{j=1}^{{\cal N}-\alpha}\left[\xi(t_{j+\alpha}) - \xi(t_{\alpha})\right]^2.~\label{MSD}
\ee
The asymptotic growth rate of the mean square displacement can be defined as
\be
K=\lim_{\alpha\rightarrow\infty}(\log M(t_{\alpha}))/\log t_{\alpha},~\label{Kdef}
\ee
which is computed, by performing a least square fit of $\log M(t_{\alpha})$ versus $\log t_{\alpha}$, in the range $1\le \alpha\le{\cal N}_1$ for a choice of ${\cal N}_1$ such that $1\ll{\cal N}_1 \ll{\cal N}$ and ${\cal N}_1={\cal N}/10$. In the definition of the asymptotic growth rate, equation\ (\ref{Kdef}), the claim is made that $\xi(t_{\alpha})$ has the diffusion properties of a Brownian-like motion when the dynamics of $\phi(t_{\alpha})$ is chaotic~\cite{Test1}. Then $K\approx0$ stands for regular dynamics and $K\approx1$ implies chaotic dynamics. To avoid possible resonances between the frequencies of the underlying dynamical system and $c$ we compute $K$ for $100$ random values of the frequency $c$ drawn from the interval ($0,2\pi$), since the test is $2\pi$-periodic in $c$. The final $K$ value is then taken as the median of the computed set~\cite{Test2}.

The functions $\xi(t_{\alpha})$ and $\eta(t_{\alpha})$ in equation~(\ref{xidef}), together with $\theta(t_{\alpha})\equiv\alpha c$, can be viewed as a component of the solution to the skew product system
\be
\eqalign{\theta(t_{\alpha+1})=\theta(t_{\alpha})+c \cr
\xi(t_{\alpha+1})=\xi(t_{\alpha})+\phi(t_{\alpha})\cos\theta(t_{\alpha}) \cr
\eta(t_{\alpha+1})=\eta(t_{\alpha})+\phi(t_{\alpha})\sin\theta(t_{\alpha}),}\label{xietadef}
\ee
driven by the dynamics of the observable $\phi(t_{\alpha})$. Here $(\theta,\xi,\eta)$ represents the coordinates on the Euclidean group of rotations $\theta$ and translations $(\xi,\eta)$ in the plane~\cite{Nicol,Field}. In~\cite{Test1} it was argued that inspection of the dynamics of the $(\xi,\eta)$-trajectories provides a quick and simple visual test of whether the underlying dynamics is regular or chaotic, and so it will be employed, along with the $K$ value, to assess the applicability of the 0-1 test to study Hamiltonian chaos.

Before continuing we have to mention that the above methodology to obtain the $K$ value is described as the \emph{regression method} in~\cite{Test3}, where a modified version of the 0-1 test, termed \emph{correlation method}, has been introduced. In the latter the $K$ value is computed as a correlation coefficient of the vectors $\{\alpha\}_{\alpha=1}^{{\cal N}_1}$ and $\{D(t_{\alpha})\}_{\alpha=1}^{{\cal N}_1}$, where $D(t_{\alpha})$ is a modified mean square displacement. We deemed unnecessary to adopt this new approach for a number of reasons. The main one is that the correlation method renders essentially the same results as the original one. The regression method (the original 0-1 test), at least for the studied cases in~\cite{Test1,Test2}, correctly classifies the studied signals. No single instance of a misclassification rendered by the regression method and later corrected by the correlation method is presented in~\cite{Test3}. Furthermore, the aforementioned comparisons are made with a short time series length (2000 data points) and only for data series corresponding to the logistic map. Now, as we will see in Secs.\ \ref{sec:TSS} and \ \ref{sec:BM}, for the herein considered systems a proper assessment of the performance of the 0-1 test can only be achieved with very long time series. But in~\cite{Test3} no results are presented wherewith the correlation method reaches the asymptotic $K$ value in a shorter time (measured by the time series length) than the regression method. The fact that the performance of both methods depends on the validity of the limit ${\cal N}_1\ll{\cal N}$ could possibly indicate that a sufficiently long time series has to be employed to correctly classify a given signal, irrespective of the employed method to obtain the $K$ value. So far there is no empirical evidence that contradicts this last statement, although we also acknowledge that this does not necessarily imply positive evidence to support it either. Nevertheless, the proper corroboration or refutation of this delicate point is out of the scope of the present work. Finally, as will be clear below, a great amount of useful information to characterize the 0-1 test can be obtained by inspecting the $(\xi,\eta)$-trajectories defined by eautions~(\ref{xietadef}), which are independent of the method, either regression or correlation, used to obtain the corresponding $K$ values. Therefore, in the rest of this work we will work exclusively with the original 0-1 test, which is simpler and has the advantage that its performance has been assessed in more cases, as already mentioned in the Introduction, than the modified one.

\section{The model\label{sec:Model}}

The Hamiltonian model we are considering can be written, in terms of dimensionless variables, as
\be
H=\sum_{i=1}^N\left[{p_i^2\over2m_i} + {1\over2}(q_{i+1}-q_i)^2 + {1\over4}\beta(q_{i+1}-q_i)^4\right] \label{HFPU},
\ee
where $\{m_i,q_i,p_i\}_{i=1}^N$ are the mass, displacement, and momentum of the $i$th oscillator, respectively, in a one-dimensional $N$ coupled anharmonic lattice; periodic boundary conditions are assumed ($q_{_{N+1}}=q_{_1}$). The value $\beta=0.1$ was used in the computation of most of the numerical results hereafter reported. Next, the $2N$ first-order Hamilton equations of motion were integrated using a third-order bilateral symplectic algorithm~\cite{Casetti}, which is a high-precision numerical scheme specially suited for long-time simulations since, with the adopted value of the rather large time step of $\Delta t=0.05$, it ensures a faithful representation of a Hamiltonian flow and keeps the total energy $E$ constant within an average fluctuation level of $\Delta E/E\approx10^{-6}$ without drift. Such a high precision in numerical integration makes the outcome of very long runs that are reported in the following reliable. Finally, from a given initial condition (to be described in the next section) we let the system evolve for about $5\times 10^6$ time steps in order to avoid any transient effects due to the chosen initial conditions before the record of the chosen dynamical variable to be studied by the 0-1 test begins.

\section{Initial conditions in the strongly chaotic regime\label{sec:IC}}

It is well known that, for high values of the total energy per degree of freedom $\epsilon\equiv E/N$, the FPU lattice is chaotic whereas, for small $\epsilon$ values, it behaves as a chain of harmonic oscillators~\cite{Pettini}, despite the presence of the anharmonic potential in the Hamiltonian~(\ref{HFPU}). Therefore, by manipulating the initial conditions, and thus the $\epsilon$ value, a very precise control of the dynamical regime in which the phase-space trajectory of the system evolves can be achieved. As a first set of initial conditions we chose, for a lattice of $N=32$ oscillators, $\{p_{_i}=0\}$ and
\be
q_{_i}=\sum_{k=1}^{N/8}\left[a_{_k}\cos\left({2\pi ki\over N}\right)+b_{_k}\sin\left({2\pi ki\over N}\right)\right].
\ee
In this way only the Fourier modes for $k\le N/8$ are different from zero at time $t=0$. The main advantage of this choice is that, since there is no randomness in these initial conditions, the obtained $K$ value will be entirely due to the intrinsic dynamics of the lattice. After choosing the $\{a_{_k},b_{_k}\}_{k=1}^{N/8}$ values and letting the system evolve for the aforementioned transient time interval, the position $\{q_{_1}(t_{\alpha})\}$ and momentum $\{p_{_1}(t_{\alpha})\}$ time series of the first oscillator in the chain were recorded. The employed sampling time $\tau=1$, which is the natural time unit, corresponds to ten times the smallest time interval available $\tau_{_\mathrm{min}}=2\Delta t=0.1$ for the employed time step~\cite{Casetti}. Furthermore, this $\tau$ value is close to the inverse of the fastest frequency of the harmonic part of equation~(\ref{HFPU}): $T_{\mathrm{min}}=2\pi/\omega_{\mathrm{max}}\equiv\pi$. Most results hereafter reported will be given in this unit of time. Finally, a time series length  of ${\cal N}=6\times10^4$ natural time units was taken.

For $\{a_{_k}=b_{_k}=3\}$, which corresponds to an energy density of $\epsilon\approx13.34$, we obtain, for the position and momentum time series, the corresponding asymptotic values for the mean square displacement $K_q=0.96$ and $K_p=0.95$ respectively. Since the $\epsilon$ value corresponds to a strongly chaotic regime, we conclude that the test is successful. In order to corroborate these results we repeated the simulations, but with the value $\beta=0$, which renders a harmonic oscillator lattice. For the choice $\{a_{_k}=b_{_k}=4.3\}$, which yields $\epsilon\approx10.3$, we obtain $K_{q}=6.1\times10^{-4}$ and $K_p=1.2\times10^{-4}$. Thus the underlying dynamics of the lattice, for these initial conditions, is well characterized by the 0-1 test.

For the next type of initial conditions we choose the equilibrium value of the oscillators displacements, i.e. $\{q_{_i}=0\}$, whereas the momenta $\{p_{_i}\}$ were drawn from a Maxwell-Boltzmann distribution at temperature $T$ consistent with a given value of the energy density, which we set as $\epsilon=10$. The random component has a self-evident physical meaning related to the impossibility of preparing any physical system in a perfectly ordered initial state: at nonzero temperature some randomness in the initial conditions is unavoidable. The obtained values are $K_q=0.91$ and $K_p=0.93$. If again the $\beta=0$ value is taken, we obtain $K_q=2.4\times10^{-3}$ and $K_p=2.6\times10^{-3}$. No significant difference whatsoever can be detected with the results of the ordered initial state. Thus we can conclude that, for the parameters so far employed, the initial conditions have no effect on the results of the 0-1 test.

\section{Weakly chaotic regime\label{sec:WCR}}

\begin{figure}
\includegraphics[width=0.98\linewidth,angle=0.0]{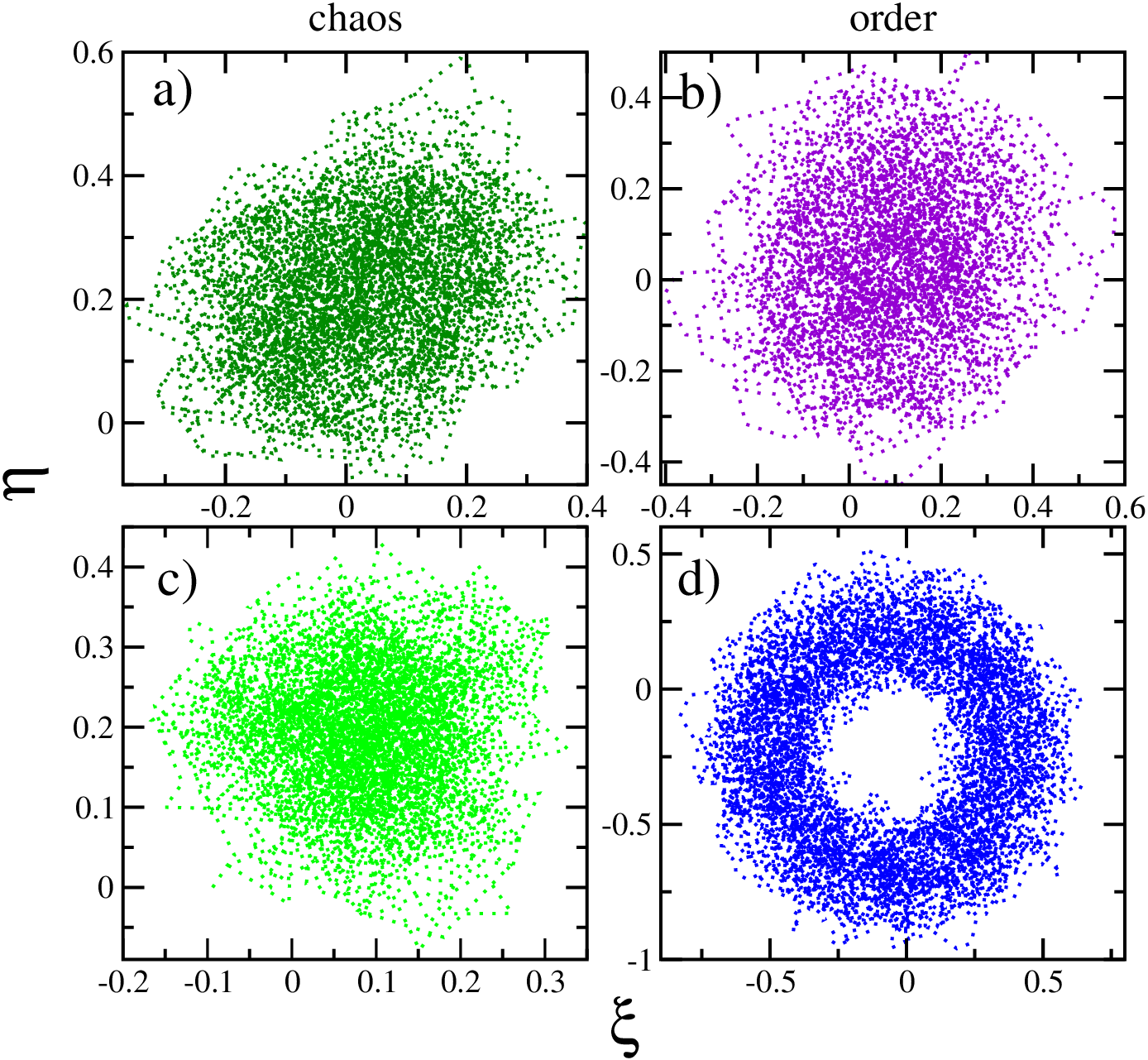}
\caption{(Color online) $(\xi,\eta)$-trajectories corresponding to the (a) position and (c) momentum time series of a single oscillator of an anharmonic FPU lattice ($\beta=0.1$). (b,d) Same as in (a,c), but for a harmonic oscillator lattice ($\beta=0$). In all cases ${\cal N}=2000$, $N=32$, and $\epsilon=0.01$ which, for the FPU lattice, corresponds to a weakly chaotic regime.}
\label{fig:e0p01}
\end{figure}

For very low $\epsilon$ values the FPU lattice is chaotic, despite the fact that, for very long times, it behaves as a harmonic oscillator chain, as already mentioned. Therefore, it is important to verify if the 0-1 test can correctly classify time records obtained in this dynamical regime. For the time series length and sampling time so far employed, the 0-1 test, with random initial conditions corresponding to $\epsilon=0.01$, yields $K_q=5.2\times10^{-3}$ and $K_p=5.0\times10^{-3}$, which implies a signal misclassification. Now, in figures\ \ref{fig:e0p01}(a) and (c) we plot, for ${\cal N}=2000$, the $(\xi,\eta)$-trajectories stemming from the position and momentum time series respectively. A stochastic-like but bounded behavior is observed in both $(\xi,\eta)$-trajectories. Henceforth we infer that diffusion in $(\xi,\eta)$ space is hindered by the low $\epsilon$ value. These results are inconclusive to unambiguously classify both signals. However, if we compare with the position and momentum $(\xi,\eta)$-trajectories resulting from a simulation with $\beta=0$ (thus rendering a harmonic chain) and plotted in figures\ \ref{fig:e0p01}(b) and (d), a visual distinction of the regular and the weakly chaotic case can be made. Therefore, the interpretation of the results in this dynamical regime could become ambiguous without the comparison afforded by the regular dynamics behavior of the momentum time series displayed in figure\ \ref{fig:e0p01}(d). Furthermore, a completely automated application of the 0-1 test, i.e. the sole reliance on the $K$ value computed from a moderate amount of data, can lead to completely wrong results if it is not properly complemented with information of the $(\xi,\eta)$-trajectory, since $K\approx0$ is obtained for all cases reported in figure\ \ref{fig:e0p01}.

The reason for the failure of the asymptotic growth rate $K$ to classify the signals corresponding to a low $\epsilon$ value can be understood in terms of the phase space structure as a function of the energy density value~\cite{Pettini}. The Hamiltonian of the FPU model can be written as
\be
H(\theta,\mathbf{I})=H_0(\mathbf{I}) + H_1(\theta,\mathbf{I}),\,\,\,\,\,\,\,\mu\equiv{\parallel H_1\parallel\over\parallel H_0\parallel}\ll1,
\ee
where ($\theta,\mathbf{I}$) are the action-angle canonically conjugated variables and $\parallel\cdots\parallel$ is a suitable norm. A consequence of the perturbation $H_1$ is that the resonant manifolds $\mathbf{n}\cdot\bomega(\mathbf{I})=0$ of $H_0$ are destroyed for any small $\mu$ and are replaced by finite-thickness chaotic layers ({\bf n} is an integer component vector and $\bomega$ is a vector whose components are $\omega_i=\partial H_0/\partial I_i$). As these chaotic surfaces intersect the constant energy hypersurface for $N\gg1$, a chaotic network (the Arnold Web) is produced which is everywhere dense in phase space. For high $\epsilon$ values the resonances are strongly overlapped and microscopic, i.e. phase space, diffusion is allowed in every direction of phase space. These facts explain the success of the 0-1 test for the $\epsilon=10$ case already studied. On the contrary, for low $\epsilon$ values resonance overlapping is drastically reduced, microscopic diffusion occurs only along resonances and thus is dramatically slowed down. These facts indeed explain both the $K\approx0$ value as well as the lack of diffusion of the $(\xi,\eta)$-trajectories for the anharmonic FPU lattice ($\beta\not=0$).

\begin{figure}
\includegraphics[width=0.98\linewidth,angle=0.0]{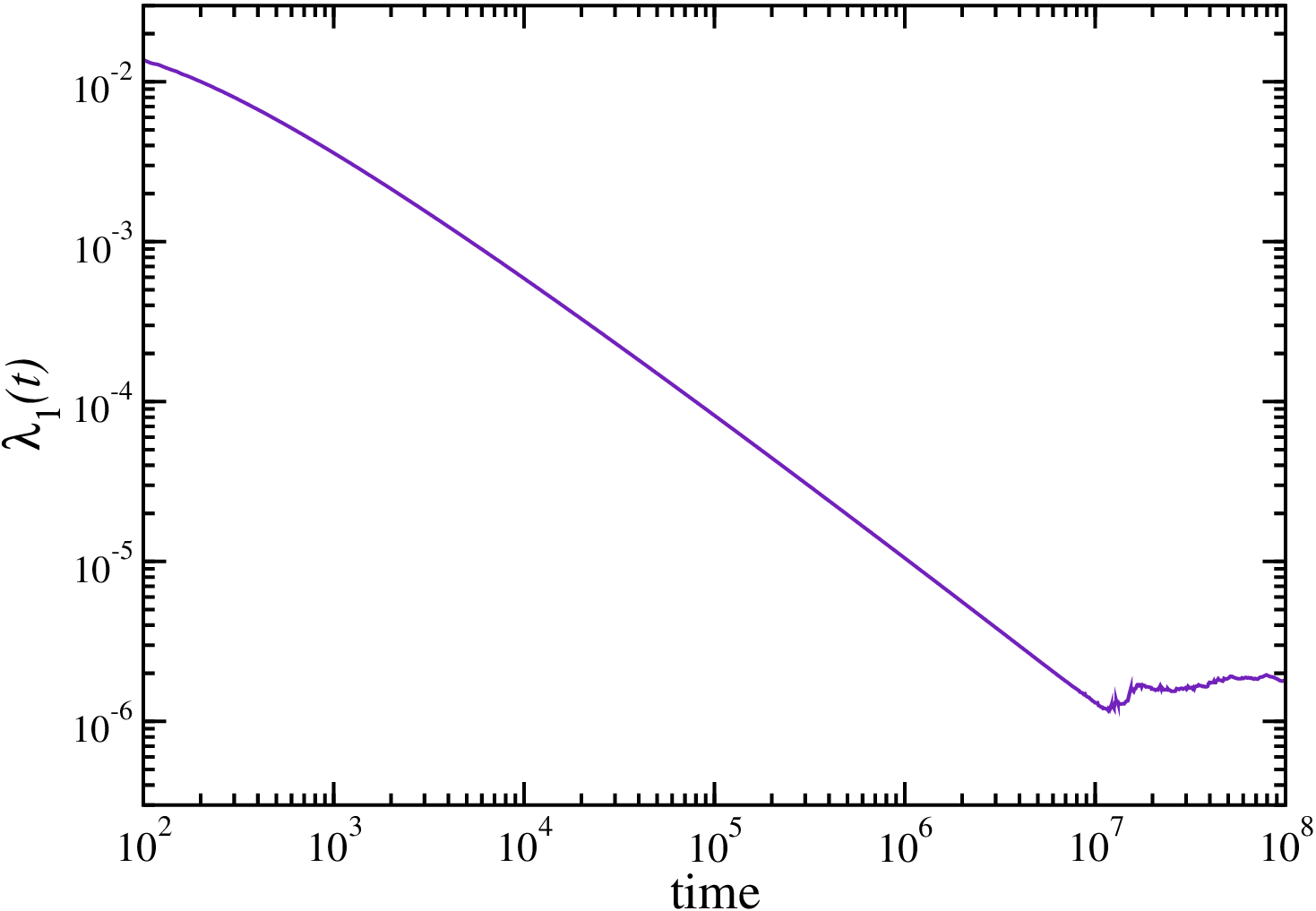}
\caption{$\lambda_1(t)$ vs time, measured in natural time units, for a FPU lattice with $N=32$ and energy density $\epsilon=0.01$.}
\label{fig:Lyap}
\end{figure}

To determine the time scale wherein the regular behavior persists, and thus a lower bound to the time series length beyond which a correct classification could be expected, we have computed the largest Lyapunov exponent (LLE) $\lambda_{_1}$ of the FPU lattice by the so called standard method~\cite{Standard}. In figure\ \ref{fig:Lyap} we report, as an example, $\lambda_{_1}(t)$ in the case $N=32$ and $\epsilon=0.01$ for random initial conditions. Up until $t\approx10^7$ the LLE seems to decay toward zero, being the behavior expected for a nonchaotic system. Then, suddenly at $t\gtrsim1.2\times10^7$, $\lambda_{_1}$ tends to converge to a nonvanishing value. This dramatic difference can be attributed to the untrapping of the FPU system from its regular region in phase space by escaping to the chaotic component of its phase space since, by the Poincar\'e-Fermi theorem~\cite{PF}, both regions are connected. Thus it is possible to define clearly what a trapping time in a regular region of phase space is; moreover, its numerical determination is unambiguous, as it can be deduced by simply looking at figure\ \ref{fig:Lyap}. Therefore it seems highly unlikely that a $K\approx1$ could be obtained with a time series length inferior to the trapping time within the phase space regular region. Indeed, for ${\cal N}=5\times10^5$, the results are $K_q=8.5\times10^{-4}$ and $K_p=1.3\times10^{-3}$. Of course, in the opposite case it can be validly inferred from figure\ \ref{fig:Lyap} that the correct classification can be obtained, but the sheer length of the required signal would render the 0-1 test impractical for the foregoing situation.

\section{Heavy impurity: time scale separation\label{sec:TSS}}

\begin{figure}
\includegraphics[width=0.98\linewidth,angle=0.0]{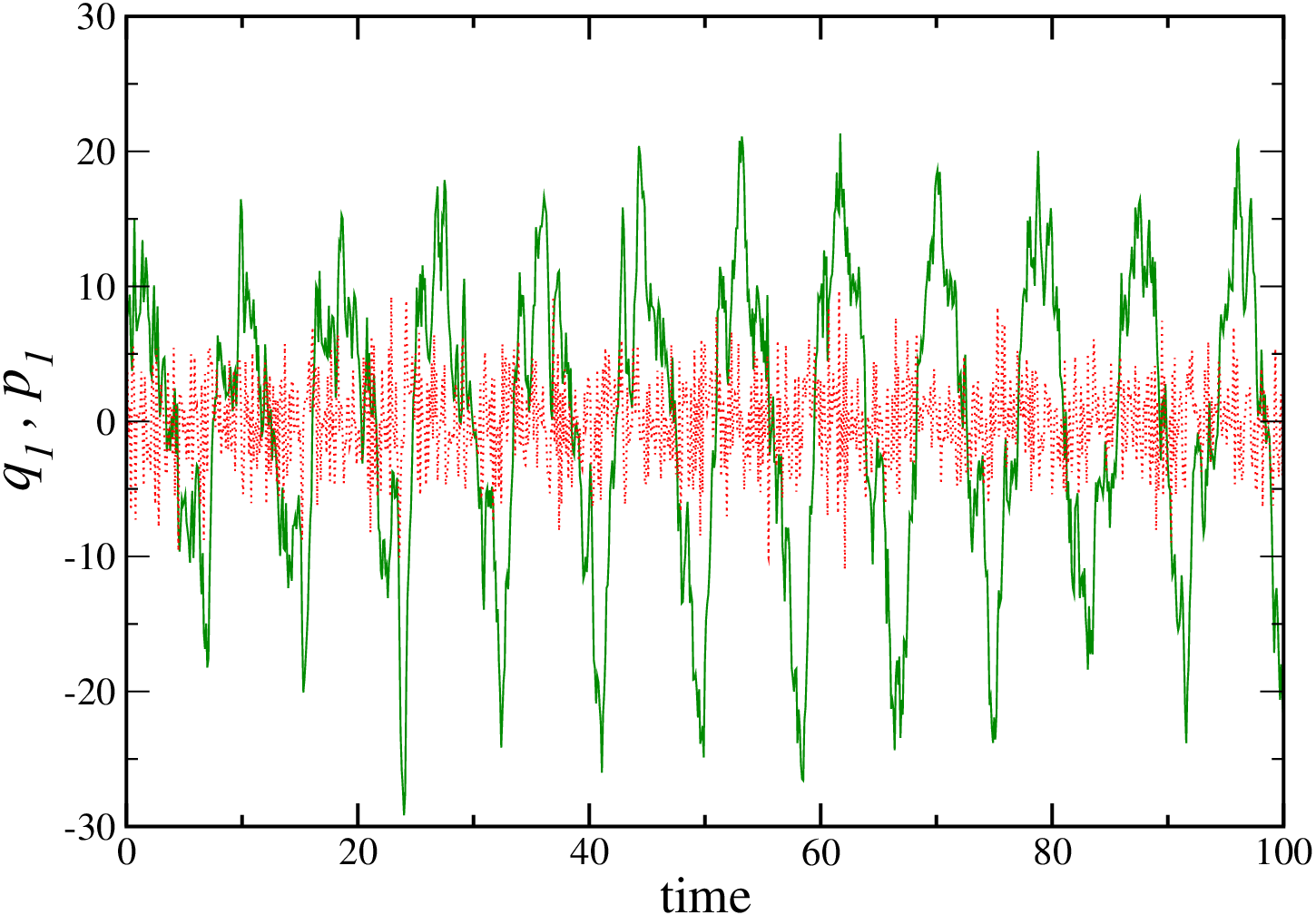}
\caption{(Color online) Position $\{q_1(t_\alpha)\}$ (solid line) and momentum $\{p_1(t_{\alpha})\}$ (dotted line) time series of the first oscillator of a lattice with $N=128$ and energy density $\epsilon=10$, which corresponds to the strongly chaotic regime. Time is measured in natural units.}
\label{fig:n128m1}
\end{figure}

From the results presented in Sec.\ref{sec:IC} it is clear that the 0-1 test can indeed classify unambiguously the dynamics of the FPU lattice in the strongly chaotic regime, i.e. for high $\epsilon$ values, irrespective of the chosen observable. The reason can be inferred from figure\ \ref{fig:n128m1}, which displays the time evolution of both the position and momentum of the first oscillator of a lattice with $N=128$ and $\epsilon=10$. It can be appreciated that the time scales in which both dynamical variables evolve are quite similar. Therefore it is reasonable to assume that both the position and momentum make an adequate sampling of the phase-space dynamics. An immediate confirmation is afforded by applying the 0-1 test, which, for ${\cal N}=6\times10^4$, yields $K_q=0.97$ and $K_p=0.92$, consistent with the known dynamical regime of the chain. 

\begin{figure}
\includegraphics[width=0.98\linewidth,angle=0.0]{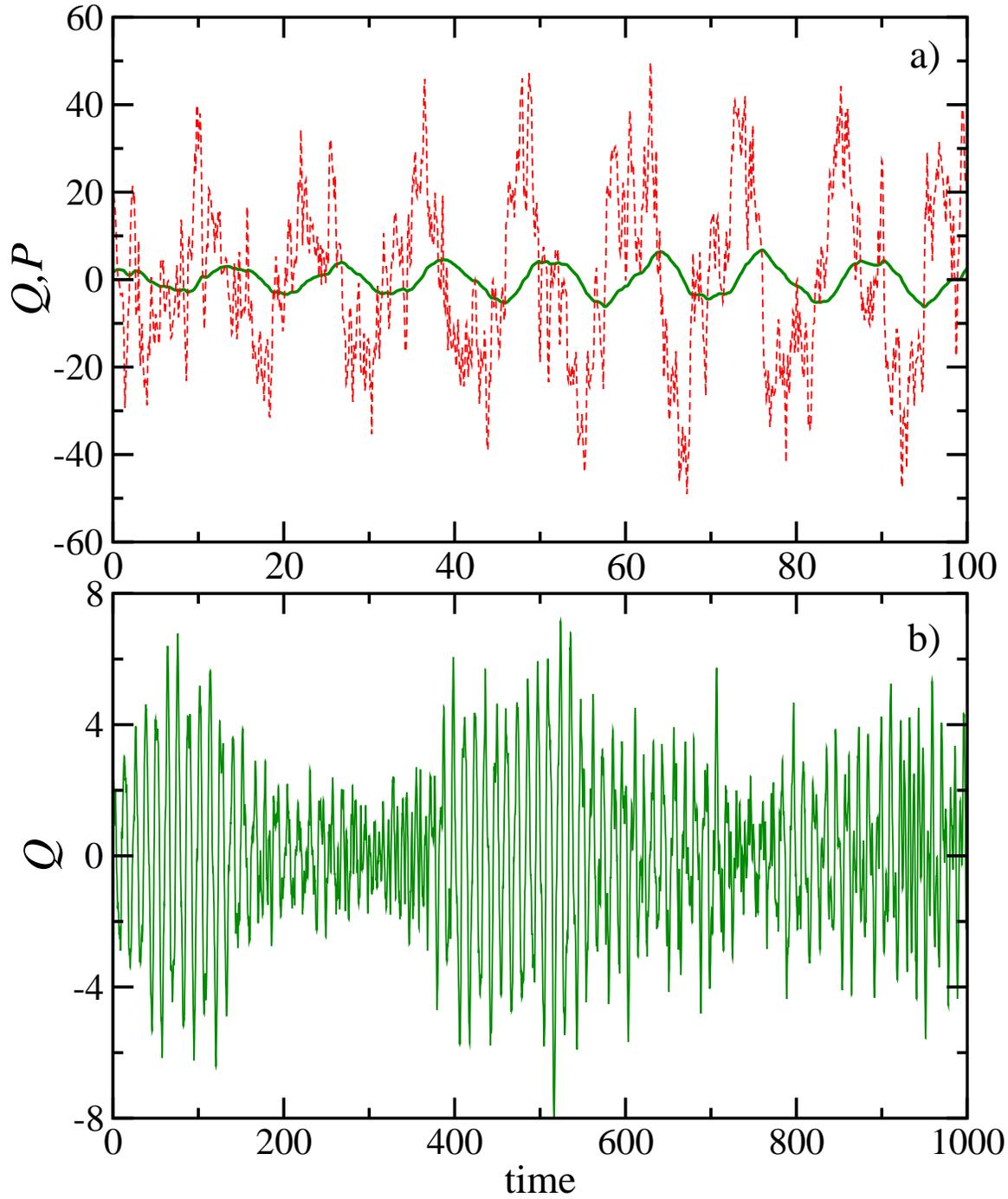}
\caption{(Color online) (a) Position $\{Q(t_\alpha)\}$ (solid line) and momentum $\{P(t_{\alpha})\}$ (dotted line) time series of a heavy impurity with $M=100$ coupled to an oscillator lattice of $N=128$ and $\epsilon=10$. (b) Same position time series as in (a), but displayed in a larger time scale. Time is measured in natural units.}
\label{fig:n128m100}
\end{figure}

A dramatic difference is obtained if the mass of the first oscillator is increased to $m_{_1}\equiv M=100$ and a lattice of $N+1$ oscillators is now taken. In figure\ \ref{fig:n128m100}(a), again for $N=128$ and $\epsilon=10$, it is clearly seen that the inertia of the heavy impurity renders the time evolution of its position $q_1\equiv Q$ quite differently to that of its conjugate momentum $p_1\equiv P$. In the displayed time interval the momentum value experiences many changes whereas the variations in the position are slower. Thus the time scales associated with the position data are much longer than the length of the data set itself, as can be seen in figure\ \ref{fig:n128m100}(b), which displays the same position time record of figure\ \ref{fig:n128m100}(a), but in a larger time scale. This time scale difference has a strong effect upon the results of the 0-1 test since, for the position time series with ${\cal N}=10^5$, a value $K_q=0.28$ is obtained, clearly inconsistent with the chaotic dynamics of the lattice. However, for the momentum $K_p=0.95$; a consistent result is obtained since it is the fast variable and thus the ``correct" observable.

Under these new conditions, for the position time series the sampling time so far employed is too small, which implies data oversampling~\cite{Test3}; an immediate solution, which is suggested by the comparison of the results in figures\ \ref{fig:n128m100}(a) and (b), is to take a coarser sampling time, and thus a longer time series length. For example, if $\tau=10$ (i.e. 10 time units) and ${\cal N}=10^5$ data points, extracted from a time series of length ${\cal N}=10^6$, are employed, the result for the position time series is $K_q=0.61$, consistent with the chaotic dynamics of the system. A more refined method to obtain the optimal sampling time is to use the first minimum of the mutual information~\cite{Kantz}. For the employed data set this method yields $\tau=30$, which is rather close to that already taken.

An observation worth making at this point is that the results of figure\ \ref{fig:n128m100} suggest that data oversampling can only be invoked as a way to correct data misclassification when there is an underlying physical mechanism responsible for it. The time series associated with the position of the heavy impurity plotted in figure\ \ref{fig:n128m100}(a) is oversampled because the time evolution of this variable occurs in a much slower time scale than that associated with the momentum of that same oscillator, as well as with the rest of the degrees of freedom of the system. However, for the weakly chaotic regime studied in Sec.\ \ref{sec:WCR} this mechanism is completely absent, a fact that renders the plots of both the position and momentum of the first (or any other) oscillator in the homogeneous, i.e. uniform mass, case (not shown) for $\epsilon=0.01$ virtually identical, except for the vertical scale, to those corresponding to the strongly chaotic regime depicted in figure\ \ref{fig:n128m1} with $\epsilon=10$. Thus the absence of a heavy impurity avoids time scale separation altogether, independently of the dynamical regime, either regular or chaotic, of the system.

However, employing a coarser sampling time, and hence a longer time series, can hardly be considered a general solution to obtain the correct $K$ value. The reason is that the feasibility to generate longer data sets to overcome the oversampling issue, and thus avoid the apparent misclassification, cannot be guaranteed in general. To address this point we will perform a detailed analysis of the $K$ dependence on both the system size $N$ and time series length ${\cal N}$ for the position of the heavy impurity and a fixed sampling time of $\tau=1$ which, as already explained, is the natural unit of time. The main reason for retaining this sampling time is that generating a long time series will \emph{always} be necessary, irrespective of the sampling time employed later on to avoid oversampling, as the aforementioned example clearly highlights. Furthermore, with $\tau=1$ and an even longer time series length, consistent values of the asymptotic growth rate can be obtained. Again for the already considered example, with ${\cal N}=6\times10^6$ we obtain $K_q=0.55$, which correctly classifies the signal.

\begin{figure}
\includegraphics[width=0.98\linewidth,angle=0.0]{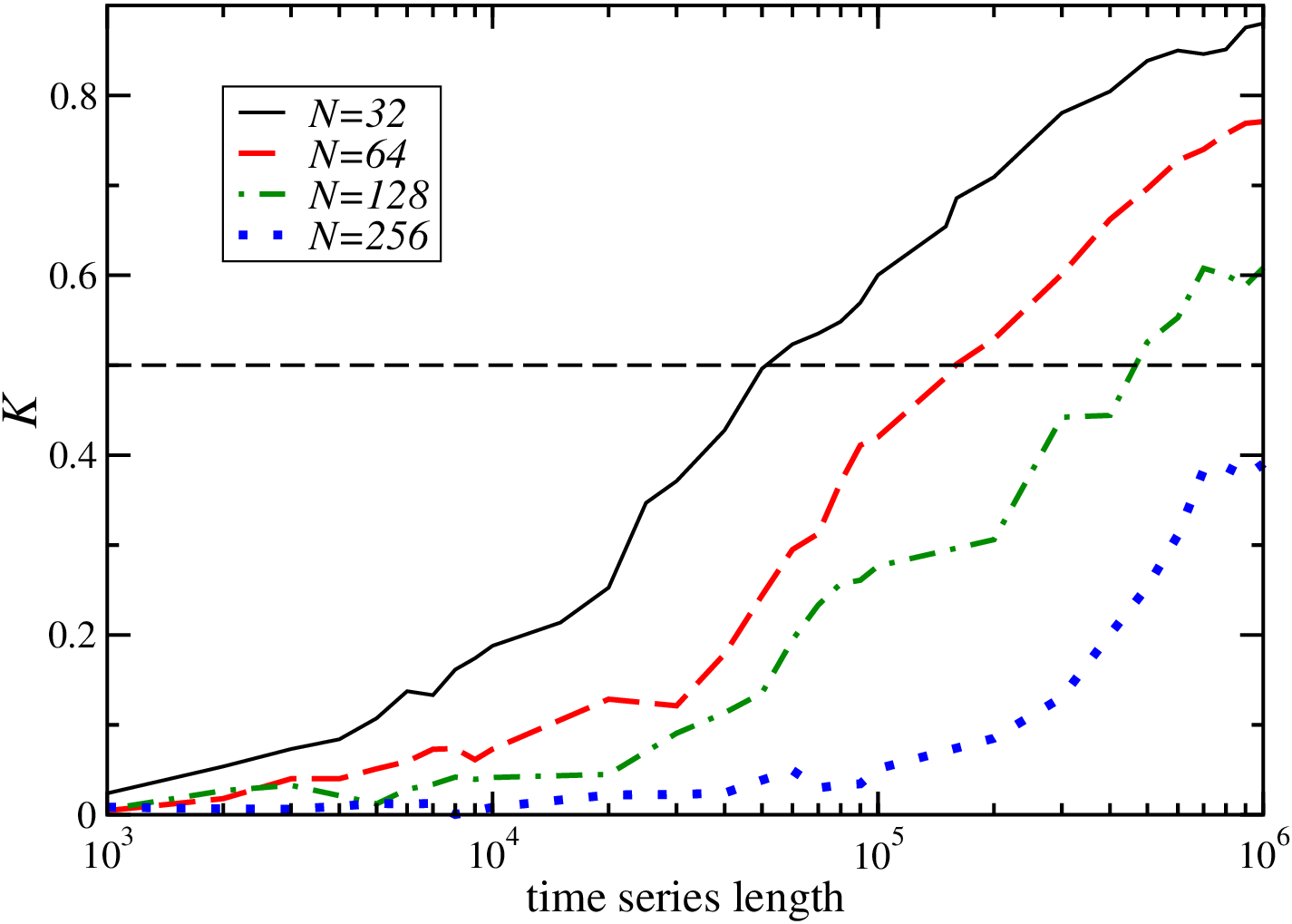}
\caption{(Color online) Asymptotic growth rate $K$ vs position time series length ${\cal N}$ for a heavy impurity with $M=100$ coupled to an anharmonic oscillator lattice with $\epsilon=10$ for various system sizes: $N=32$ (solid line), $64$ (dashed line), $128$ (dashed-dot line), and $256$ (dotted line). The horizontal dashed line indicates the $K=0.5$ value.}
\label{fig:KvsTSL}
\end{figure}

Figure\ \ref{fig:KvsTSL} presents the results of the dependence of $K$ on ${\cal N}$ for various $N$ values. As can be appreciated, for small lengths the 0-1 test yields $K_q\approx0$, irrespective of the oscillator number $N$. However, as the length of the employed time series is increased, $K_q\rightarrow1$, albeit at a rate that rapidly diminishes as the system size $N$ increases, thus making an automated application of the test seemingly unfeasible. Although it is also clear from the figure that, if a sufficiently large time series is employed, the correct $K$ value can always be obtained for this case, this option becomes increasingly impractical as $N$ increases. However, to explore the possibility of an automated application, notwithstanding the information rendered by visual inspection of the plot $K$ vs ${\cal N}$ in figure\ \ref{fig:KvsTSL}, it is necessary to determine the $N$ dependence of the critical time series length ${\cal N}$ beyond which $K$ correctly classifies the signal. Such dependence has to be weak enough to allow the possibility of an automated application of the 0-1 test.

\begin{figure}
\includegraphics[width=0.98\linewidth,angle=0.0]{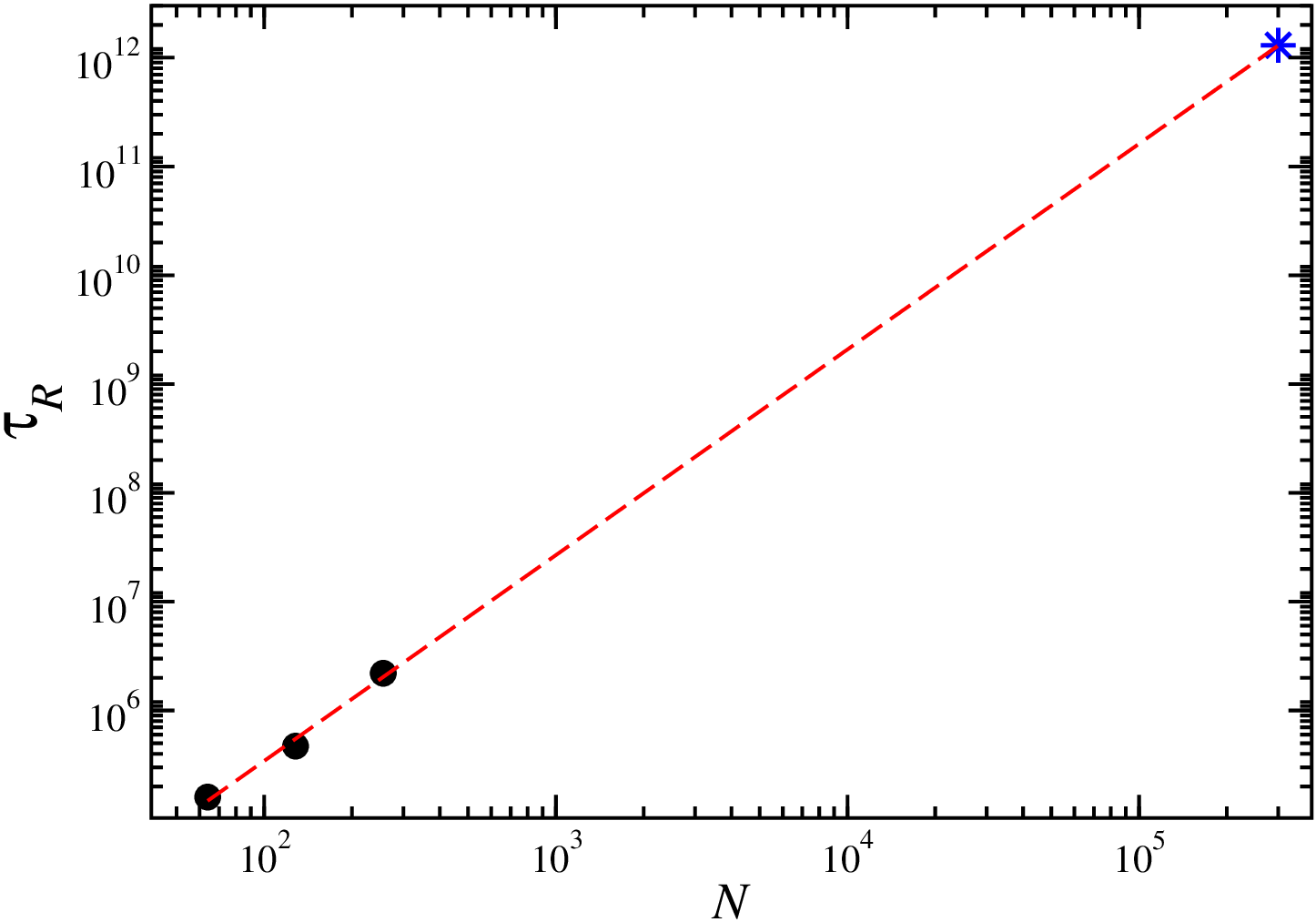}
\caption{(Color online) Time series length $\tau_{_R}$ required for $K$, computed from the position time series of a heavy impurity of $M=100$ coupled to an anharmonic oscillator lattice, to reach the value $0.5$ vs system size $N$ with $\epsilon=10$. The asterisk represents the extrapolation of the relaxation time to the case $N=300\,000$.}
\label{fig:RTvsN}
\end{figure}

It is clear from figure\ \ref{fig:KvsTSL} that the time series length ${\cal N}$ required to obtain a correct classification of the signal grows as the system size $N$ increases. To obtain an unambiguous $K$ value that could make an automated application of the test feasible, a lower bound to the asymptotic growth rate $K$ has to be established. If we consider that $K=0.5$ is the minimum value needed to unambiguously consider the underlying dynamics as chaotic, then we can define $\tau_{_R}$ as the time series length needed to attain the aforementioned $K$ value. In figure\ \ref{fig:RTvsN}, $\tau_{_R}$ is plotted as a function of the system size for the $N$ values depicted in figure\ \ref{fig:KvsTSL}. We obtain a strong system size dependence of the form $\tau_{_R}\sim N^{1.9}$, a result that indicates the unfeasibility to obtain data for larger system sizes. The aforementioned scaling, although not highly accurate due to the small data set employed, is nevertheless rendered plausible by the systematic behavior of $K$ as a function of $N$ depicted in figure\ \ref{fig:KvsTSL}. Furthermore, there is no reason to believe that the already noticed tendency will be modified for larger $N$ values. If we extrapolate the relaxation time to the case of a lattice of $N=300\,000$ oscillators, we find $\tau_{_R}\approx1.3\times10^{12}$ (as indicated by an asterisk in that same figure). Therefore, for very large lattices the position time series would have to be extremely large (and prohibitively expensive to compute) in order to classify the signal as chaotic by means of the 0-1 test with the above defined criterion.

\begin{figure}
\includegraphics[width=0.98\linewidth,angle=0.0]{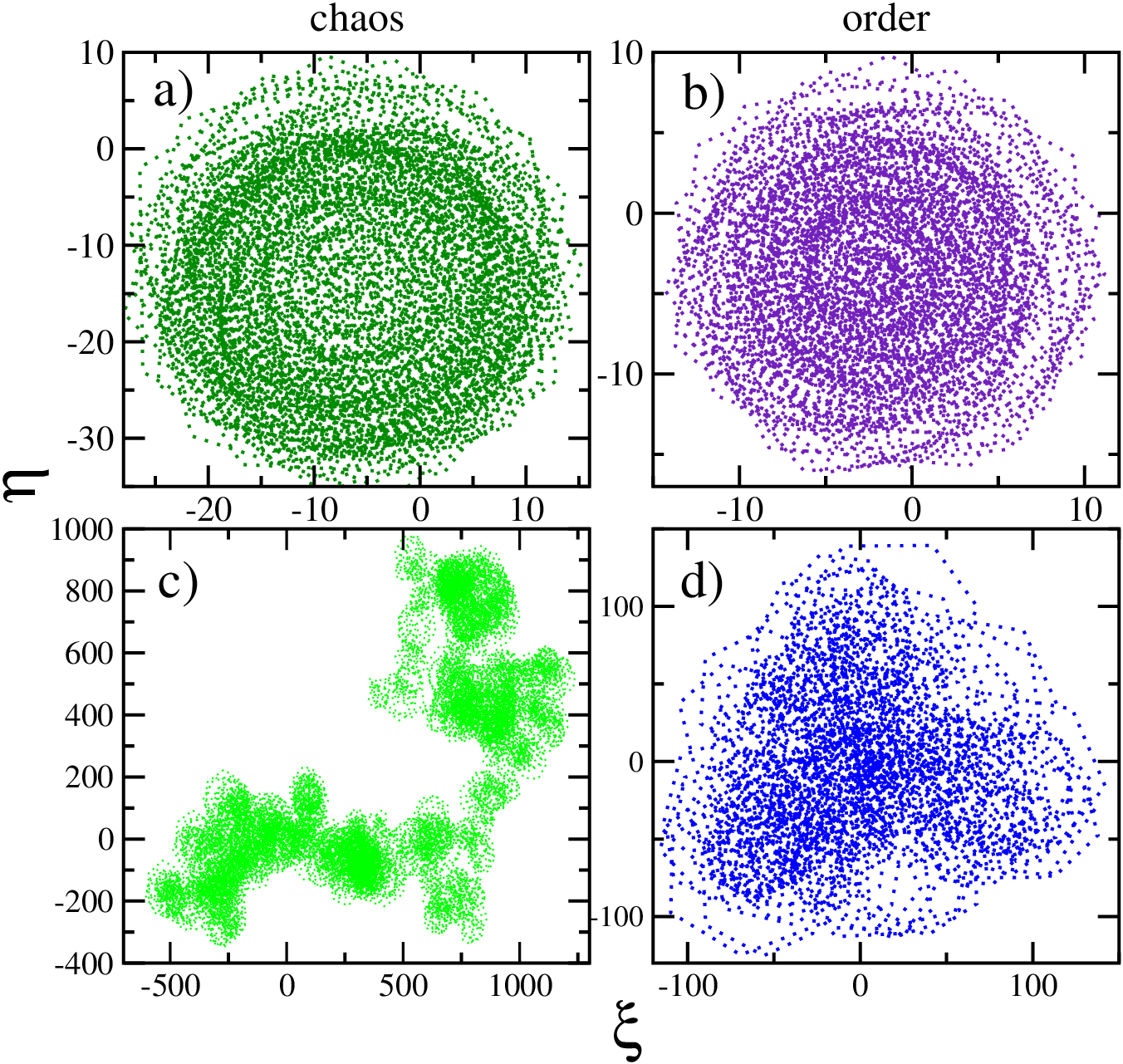}
\caption{(Color online) $(\xi,\eta)$-trajectory corresponding to the heavy impurity (a) position and (c) momentum time series with a value of $\beta=0.1$ in both instances. (b,d) Same as in (a,c), but for a harmonic oscillator lattice ($\beta=0$). $M=100$, $N=256$, $\epsilon=10$, and ${\cal N}=2\times10^4$ (although a smaller number of points is displayed for clarity) in all cases.}
\label{fig:N256}
\end{figure}

Since for the case of a low $\epsilon$ value and a homogeneous (uniform mass) lattice a simple inspection of the $(\xi,\eta)$-trajectory was helpful to classify the considered signal, we proceed to corroborate if this strategy remains useful for the new conditions under study. In figure\ \ref{fig:N256} we present the $(\xi,\eta)$-trajectories, for ${\cal N}=2\times10^4$, corresponding to the (a) position and (c) momentum time series of the heavy impurity for a lattice with $N=256$, $\epsilon=10$, and $M=100$. In the first case the slow variable yields an apparent regularity, whereas in the second the obtained unbounded and diffusive-like behavior is a clear signature of the underlying chaotic dynamics. In figures\ \ref{fig:N256}(b) and (d) we plot the same $(\xi,\eta)$-trajectories, but for $\beta=0$, which correspond to a harmonic lattice. The behavior for the position time series is indistinguishable from the corresponding behavior in the chaotic regime. For the momentum time series, the situation is drastically different: no unbounded, diffusive-like behavior is observed whatsoever. Thus the test yields a misclassification (due to the short length of the employed time series, as inferred from the results of the last paragraph) of the position signal stemming from a chaotic dynamics, whereas for the corresponding momentum time series a correct classification is obtained both by direct inspection of the $(\xi,\eta)$-trajectory and with the asymptotic growth rate, since $K_p=0.83$ in this case ($K\approx0$ in all other instances). Our next objective will be to corroborate if the aforementioned results remain valid for the largest system size considered: $N=300\,000$. 

\section{Large system size limit and brownian motion\label{sec:BM}}

\begin{figure}
\includegraphics[width=0.98\linewidth,angle=0.0]{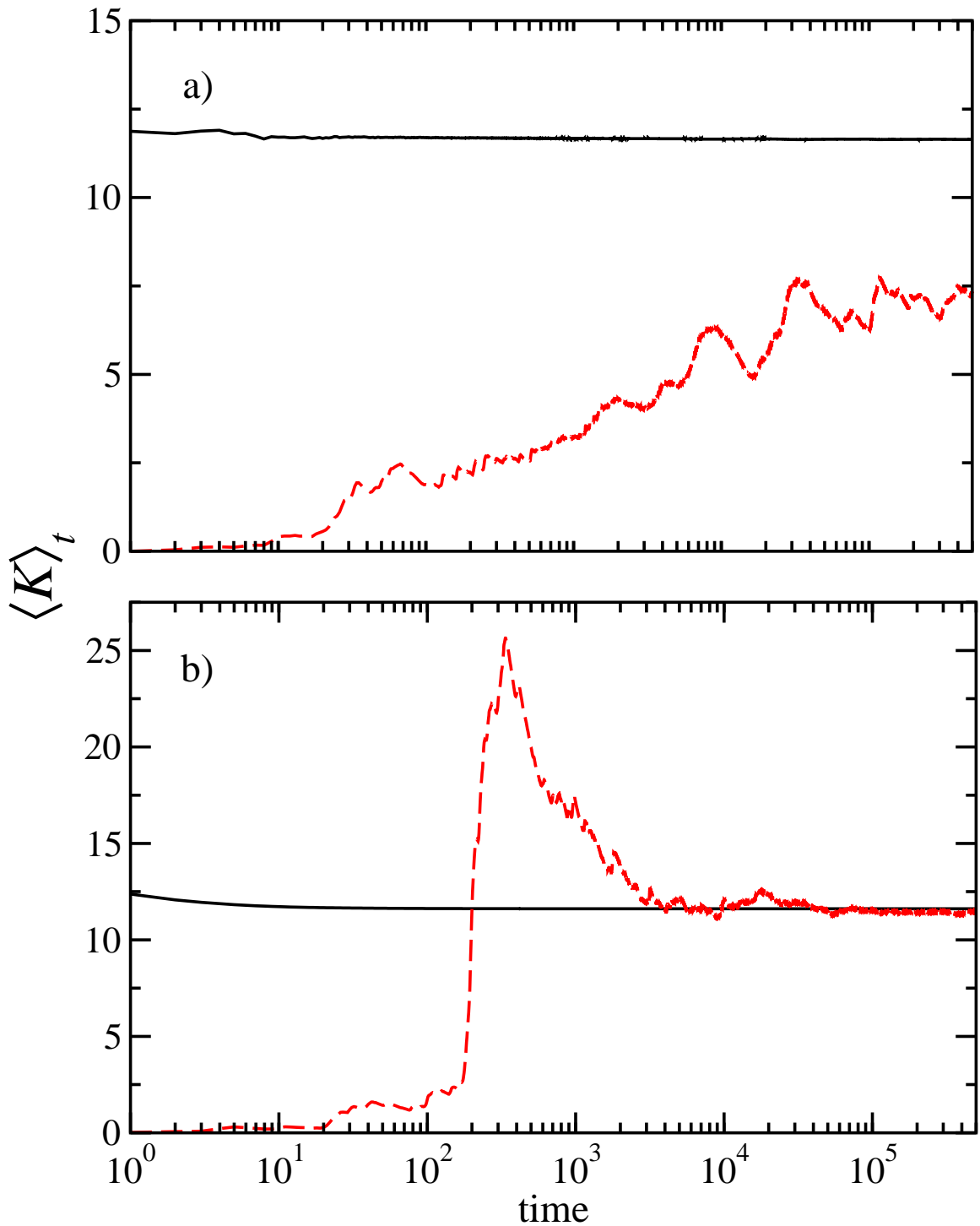}
\caption{(Color online) (a) Average temperature (continuous line) of a FPU oscillator lattice with $N=128$ and average kinetic energy (dashed line) of a coupled heavy impurity with $M=100$ vs time for an energy density $\epsilon=10$. (b) Same information as in (a), but for a lattice with $N=300\,000$ oscillators. Time is measured in natural units.}
\label{fig:SmallBig}
\end{figure}

If a heavy impurity is embedded in the oscillator lattice, large $N$ values are unavoidable for a number of reasons. The first one is that only in this limit the heavy impurity and the oscillator lattice are in thermal equilibrium. To meet this condition the mean kinetic energy $\langle K_{_{BP}}\rangle_t\equiv\langle P^2/M\rangle_t$ of the heavy impurity (where $\langle\cdots\rangle_t$ means temporal average) has to be approximately equal to the mean temperature $T_{cin}\equiv\langle\sum_{i=1}^N p_i^2\rangle_t$ of the oscillator chain, which plays the role of a thermal bath. In figure\ \ref{fig:SmallBig} we present the time evolution of the aforementioned variables for large and small lattices starting from the random initial conditions described in Sec.\ \ref{sec:Model} during the time interval before the recording of the position and momentum time series begins. It can be observed that only in the case of a large lattice the thermal equilibrium within the depicted time scale is properly established, whereas for the small lattice a metastable, non-thermodynamic state is reached.

The second and more important reason for taking large $N$ values is because, after thermal equilibrium is reached, it has been explicitly shown that, for the $\epsilon$ values so far considered and $N=300\,000$, the heavy impurity performs Brownian motion~\cite{RomeroBraun} and can thus be rightly termed BP. Furthermore, it has also been shown that the dynamics of the lattice is not affected by the presence of the impurity~\cite{Romero}. Finally, it has also been shown that the transition between weak and strong chaos can be detected by applying the standard techniques of nonlinear time series analysis of~\cite{Kantz} to the momentum time series of a heavy impurity coupled to a FPU lattice of $N=300\,000$ light oscillators~\cite{RomeroDracuBraun}. Henceforth the 0-1 test will be applied to a BP position time record of length ${\cal N}=2\times10^5$ after equilibration.

From the above presented evidence it is clear that this one dimensional microscopic model is the simplest one that captures the essential details of the full three-dimensional Brownian motion in fluids. Therefore it is physically meaningful to compare the results obtained by applying the 0-1 test to the position time series of the BP of this simple model to those of actual experimental records. The most precise available data are those obtained in 1998 by P. Gaspard {\sl et al}.~\cite{Gaspard} from the observation of the quasi two-dimensional Brownian motion of a colloidal particle, which has a diameter of $2.5\mu$m, suspended in deionized water at $22^{\circ}$C. In this case the time series $\{Q(t_{\alpha})\}$ corresponds to the time record of the $x$ component of the position, measured in $\mu$m, of the colloidal particle with a sampling time of $\tau=1/60$ s and ${\cal N}=145\,612$. See~\cite{Biggs} for further experimental details.

\begin{figure}
\includegraphics[width=0.98\linewidth,angle=0.0]{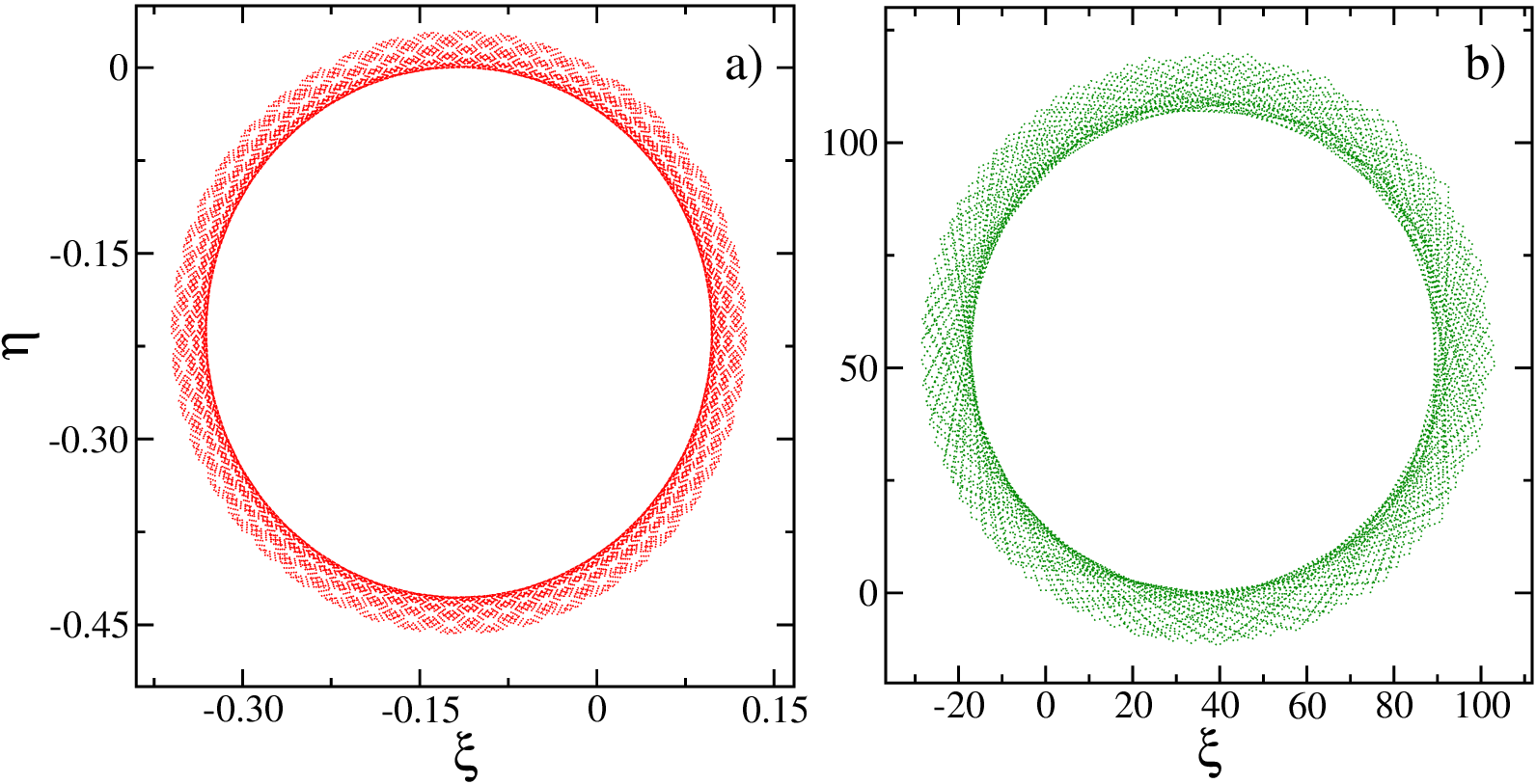}
\caption{(Color online) (a) $(\xi,\eta)$-trajectory obtained from the position time record $\{Q(t_{\alpha})\}$, with $\tau=1$ and ${\cal N}=2\times10^5$, of a heavy impurity with $M=100$ embedded in an anharmonic lattice of $N=300\,000$ oscillators. (b) $(\xi,\eta)$-trajectory obtained from the experimental time record, with $\tau=1/60$ s and ${\cal N}=145\,612$, of the position of a colloidal particle suspended in water.}
\label{fig:PosAE}
\end{figure}

The $(\xi,\eta)$-trajectories obtained from the artificial and experimental time series are displayed is figures\ \ref{fig:PosAE}(a) and (b), with computed values of $K=3\times10^{-3}$ and $K=2\times10^{-2}$ respectively. As can be readily appreciated, these results seem to indicate a periodic, non ergodic dynamics underlying both time series. For the artificial time series this result seems at odds with the information available from the LLE, $\lambda_{_1}=0.12$ for $\epsilon=10$, which clearly indicates that the system is strongly chaotic~\cite{Pettini,Romero}. In the case of the experimental data, the power spectrum $P(\omega)\sim\omega^{-2}$ indicates that the motion is of Brownian (stochastic) character~\cite{Gaspard}. Therefore, an apparent misclassification of both types of series is obtained.

However, in view of the results already presented in figures\ \ref{fig:n128m100} and\ \ref{fig:RTvsN} the described phenomenology can be attributed to a finite-size effect that has its origin in the physical issue of time scales ---the characteristic time scale of a BP's position is vastly greater than that corresponding to its momentum, which evolves in a much shorter, i.e. faster, time scale~\cite{RomeroBraun}. In fact, for the type of data of which the employed experimental time series is a representative example, it has been estimated that the necessary number of data points to detect an underlying dynamics has to be at least $\sim10^{34}$~\cite{Dettmann}. An indirect confirmation of this estimation is afforded by the additional fact that, for both artificial an experimental time series, no minimum can be identified in the mutual information, which can be considered as evidence that the ``correct" sampling time is indeed much greater than the length of the time series themselves. Therefore the results of the 0-1 test are consistent with those of nonlinear time series analysis, which are incapable, due also to finite-size effects, to render conclusive evidence of the microscopic chaos of the thermal bath wherein the BP is embedded.

\begin{figure}
\includegraphics[width=0.88\linewidth,angle=0.0]{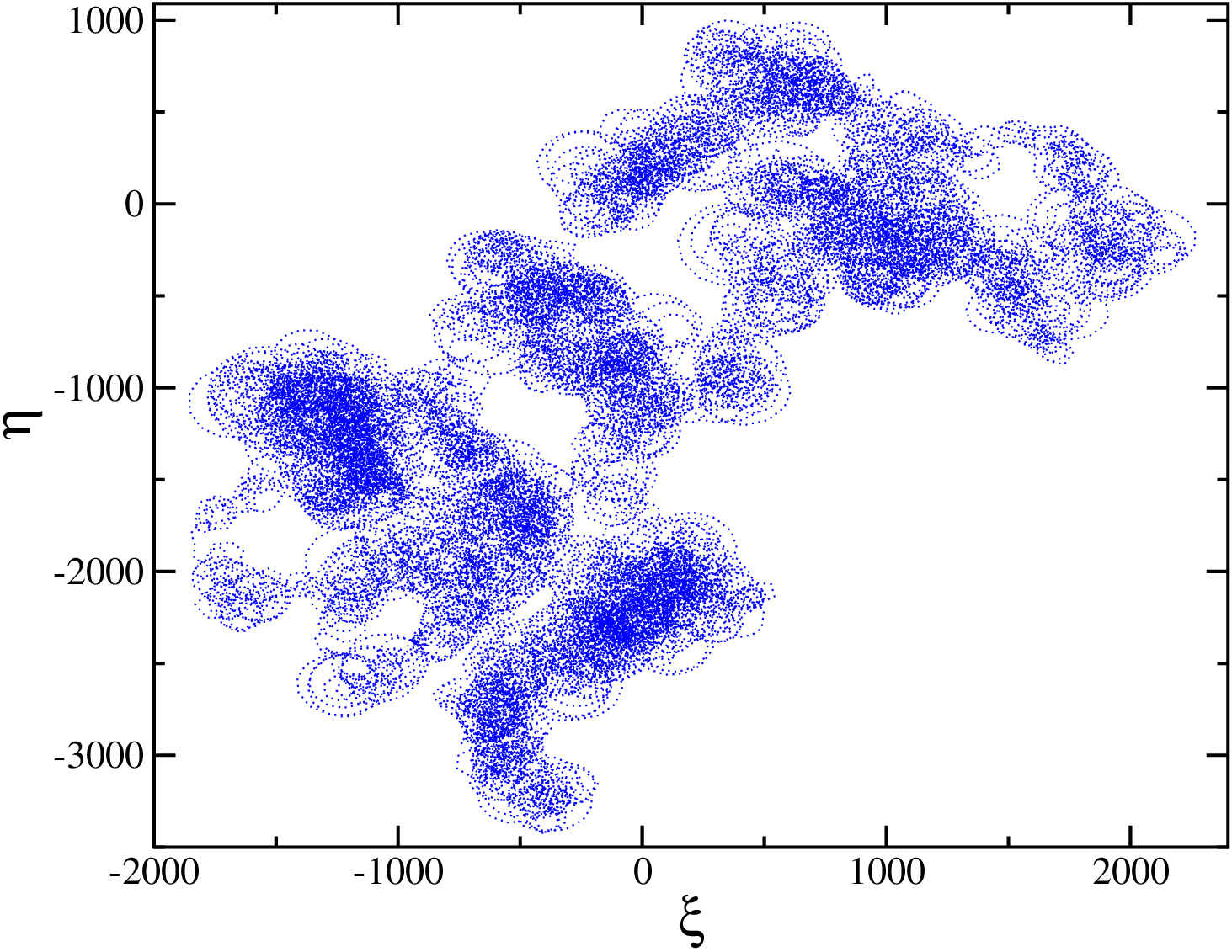}
\caption{$(\xi,\eta)$-trajectory obtained from the momentum time record $\{P(t_{\alpha})\}$, with $\tau=1$ and ${\cal N}=2\times10^5$, of a heavy impurity with $M=100$ embedded in an anharmonic FPU lattice of $N=300\,000$ oscillators for strong chaos, i.e. $\epsilon=10$.}
\label{fig:MomBP}
\end{figure}

\begin{figure}
\includegraphics[width=0.88\linewidth,angle=0.0]{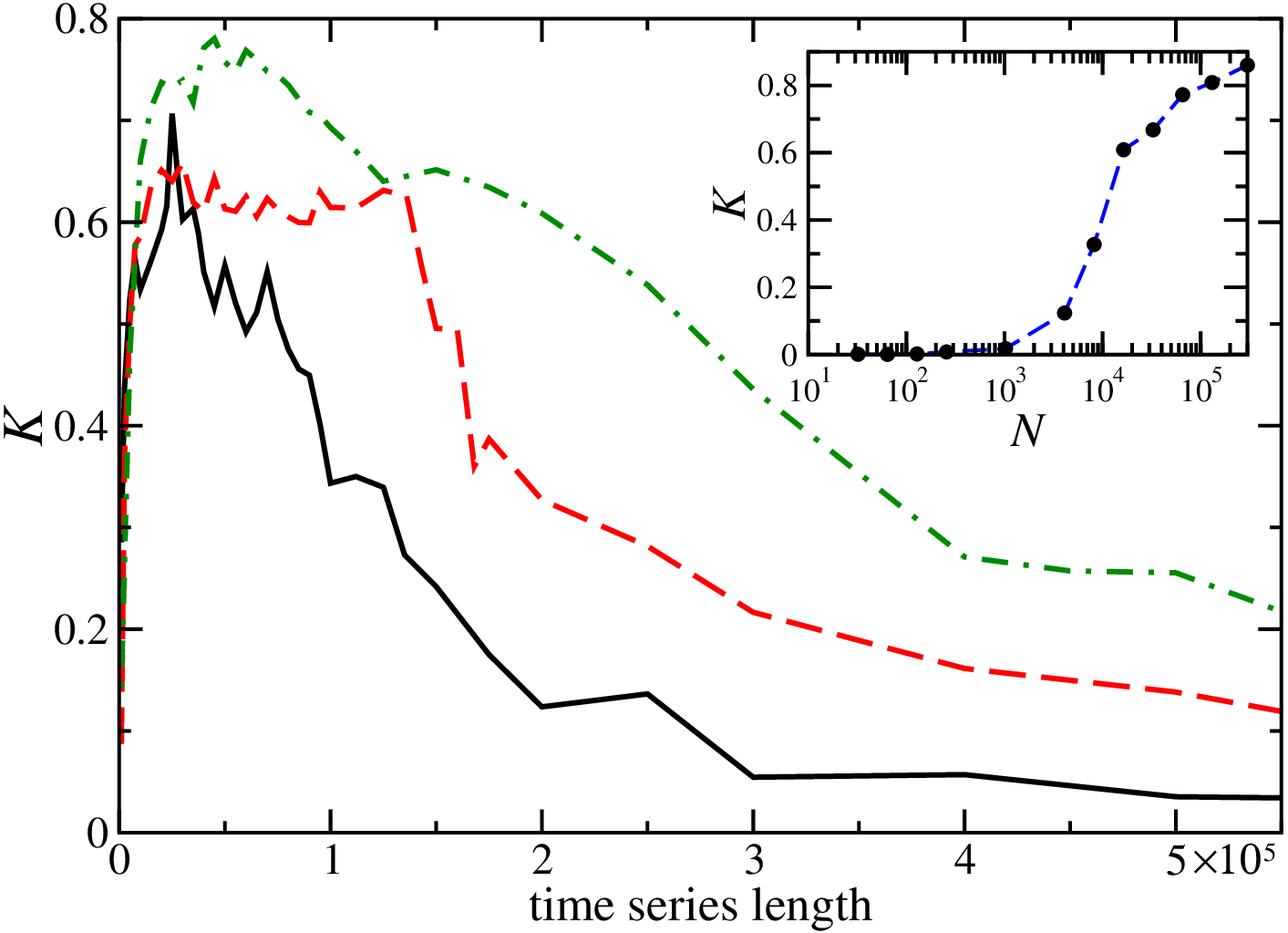}
\caption{Asymptotic growth rate $K$ vs momentum time series length ${\cal N}$ of a heavy impurity embedded in a harmonic lattice for a system size of $N=4096$ (continuous line), $N=8192$ (dashed line), and $N=16384$ (dashed-dot line). Same $M$, $\epsilon$, and $\tau$ values as in figure\ \ref{fig:MomBP}. The inset displays, for the same system, $K$ vs $N$ for a fixed time series length of ${\cal N}=2\times10^5$.}
\label{fig:KvsTSLP}
\end{figure}

For the momentum time series of the BP coupled to a FPU lattice with $N=300\,000$ and $\epsilon=10$, the 0-1 test yields $K_p=0.96$, with an unbounded, stochastic-like behavior of the ($\xi,\eta$)-trajectory, as can be seen in figure\ \ref{fig:MomBP}. A seemingly correct classification is obtained, in apparent agreement with the results of figure\ \ref{fig:N256} and the corresponding $K$ values for $N=256$. However, if a harmonic lattice is taken instead ($\beta=0$), again with $N=300\,000$ and $\epsilon=10$, the result is $K_p=0.86$, with a corresponding ($\xi,\eta$)-trajectory (not shown) virtually identical to that displayed in figure\ \ref{fig:MomBP} for the anharmonic FPU lattice. To explain this seemingly odd outcome of the 0-1 test in figure\ \ref{fig:KvsTSLP} we present the results of the dependence of $K_p$ on ${\cal N}$ for various $N$ values. It is clear that, for extremely short times, the 0-1 test detects the stochasticity of the initial conditions (which the test identifies as dynamical chaos; recall that it can not distinguish between chaos and stochastic dynamics~\cite{Test1,Reliability,Test2}), whereas for large times the correct $K_p$ value is obtained. However, it is also clear that, as the system size grows, the time scale wherein the 0-1 test yields $K_p\approx1$ also increases. With a similar extrapolation to that performed in figure\ \ref{fig:RTvsN} we obtain a times series length value of $3.8\times10^6$ for the 0-1 test to yield a value $K\approx0.5$ with $N=300\,000$, which is larger by an order of magnitude than the employed time series length of ${\cal N}=2\times10^5$, being the latter rather close to the length of the experimental time series. Thus, a misclassification is expected for a short time series length such as that currently being employed. Indeed, in the inset of the same figure it is clear that, for this fixed ${\cal N}$ value (larger values become increasingly impractical to be obtained as the system size approaches $N=300\,000$), the $K_p$ value grows steadily, from $K_p\approx0$, to 1 as the system size $N$ increases.

\section{Discussion and conclusions\label{sec:Disc}}

The first assertion that can be made from our results is that the value of the asymptotic growth rate $K$, computed from a data set of moderate length, is not in general a reliable indicator of the underlying dynamics of the FPU lattice. For the homogeneous (uniform mass) case with a low $\epsilon$ value, i.e. weakly chaotic regime, $K\approx0$ (with $\beta=0.1$), whereas the LLE is $\lambda_1\approx1.8\times10^{-6}\neq0$. In the case of the momentum time series of a heavy impurity coupled to a harmonic lattice ($\beta=0$) $K\approx1$. Thus we have provided two explicit examples for which the variable $K$ erroneously classifies the considered signals. These results seem to suggest that the systems studied in~\cite{Test1} were not complex enough to highlight the limitation of $K$ as a proper classifying variable that stems from our results obtained with the FPU lattice.

It was explicitly mentioned in~\cite{Test1}, and further stressed in~\cite{Reliability,Test2,Test3}, that visual inspection of the plot in the $(\xi,\eta)$ plane is effective to distinguish between regular dynamics and chaos. Our results in figure\ \ref{fig:e0p01}(d), where a regular and bounded $(\xi,\eta)$-trajectory is obtained for a harmonic chain, and those in figures\ \ref{fig:N256}(c) and \ \ref{fig:MomBP}, where an unbounded and diffusive-like behavior is present for an anharmonic FPU lattice, seem to support the aforementioned claim. However, it has to be stressed that, for the cases depicted in figures\ \ref{fig:e0p01}(a,c), the correct classification could be obtained only because there was additional information available, namely the results in figures\ \ref{fig:e0p01}(b,d), to make the crucial comparison between two different dynamical regimes. With just the results of figures\ \ref{fig:e0p01}(a,c), and no information whatsoever about their origin, there is no way to determine the dynamical regime corresponding to each one. More explicitly, each of the plots in figures\ \ref{fig:e0p01}(a) and (b) for position time series could be interpreted as stemming from a regular dynamics, even though the first one corresponds to a weakly chaotic regime. Only comparing the results for the momentum time series, figures\ \ref{fig:e0p01}(c,d), could the proper classification be performed. Finally, the diffusive-like behavior displayed in figure\ \ref{fig:MomBP} for the anharmonic FPU lattice was also obtained for the case of a harmonic lattice, being the result of the insufficient time series length employed in the latter case, as inferred from figure\ \ref{fig:KvsTSLP}. Thus, within the time scales studied, the 0-1 test renders inconclusive results for all considered situations. Furthermore, it can be said that, in general, there is no guarantee that the test works without additional information concerning the considered system. 

Nevertheless, it could be argued that all of the above problems can always be solved by taking a longer time series length, since, as argued in~\cite{Test1} and~\cite{Test3}, from the results of~\cite{Nicol} and~\cite{Melbourne} it follows that, in principle, the 0-1 test works with probability one as ${\cal N}\rightarrow\infty$. However, in any practical situation, such as the one currently being addressed, only a finite number of data points are available (a situation especially clear for experimental series) and thus the issue of time scales wherein the 0-1 test is valid becomes unavoidable. The estimation of the required times to obtain the correct $K$ value for a system size of $N=300\,000$, inferred from figure\ \ref{fig:RTvsN} and\ \ref{fig:KvsTSLP}, are large enough to clearly render the 0-1 test impractical for the considered setup. This situation is specially unsettling since, for that very same system size $N$, the LLE has been computed for the $\epsilon$ values herein considered~\cite{Romero}.

From the above discussion it would not be entirely correct to infer that the 0-1 test is invalid; the most appropriate conclusion to be drawn from our results would be that the test has some important limitations that were not previously noticed and that reduce its range of applicability. Furthermore, we can conclude that, due to its inefficiency in probing the Hamiltonian chaos of the FPU lattice, in general the 0-1 test is not an useful tool for exploratory purposes in the case of data with no \emph{a priori} knowledge of the underlying dynamics. Nevertheless, it is also important to stress that the main limitation of the 0-1 test herein highlighted, i.e. its difficulty to cope with chaos detection (specially in the weakly chaotic regime) for signals of limited length, is not specific to the 0-1 test, but is an inherent problem of time series methods in general. Our results only suggest that, for systems in the weakly chaotic regime or with dissimilar time scales, its application is impractical. However, if the question is posed as to whether, in any other situation different to the aforementioned ones, the 0-1 test can indeed outperform traditional phase space reconstruction methods or not, we believe, based on the herein presented analysis, that the answer can only be provided on a case-by-case basis.
 
\ack
We are grateful to Matt~Briggs for access to the employed experimental time series. One of the authors (M. R. B.) wishes to thank M C Nu\~nez-Santiago and M S Romero-Nu\~nez for their comments and suggestions. Financial support from CONACyT, M\'exico is also acknowledged.




\end{document}